\pdfoutput=1 
\documentclass[aps,prl,reprint,modulo,showpacs,groupedaddress,amssymb]{revtex4-1}
\usepackage{lineno}
\modulolinenumbers[1]
\usepackage{preprintcover}
\usepackage[pdftex]{graphicx}
\usepackage{atlasphysics}
\PassOptionsToPackage{hyphens}{url}
\usepackage{xspace}
\usepackage{hyperref}
\usepackage{multirow}
\usepackage{placeins}
\usepackage{siunitx}

\newcommand*{\numRF}[2]{\num[round-mode=figures,round-precision=#2]{#1}}
\newcommand*{\numRP}[2]{\num[round-mode=places, round-precision=#2]{#1}}

\def\antibar#1{\ensuremath{#1\bar{#1}}}

\def\ttbar{\antibar{t}}

\def\pt{\ensuremath{p_{\mathrm{T}}}} % Subscript roman not italic (EE)
\def\pT{\ensuremath{p_{\mathrm{T}}}} % Subscript roman not italic (EE)
\def\met{\ensuremath{E_{\mathrm{T}}^{\mathrm{miss}}}} % Sub/superscript roman not italic (EE)

\def\TeV{\ifmmode {\mathrm{\ Te\kern -0.1em V}}\else
                   \textrm{Te\kern -0.1em V}\fi}%
\def\GeV{\ifmmode {\mathrm{\ Ge\kern -0.1em V}}\else
                   \textrm{Ge\kern -0.1em V}\fi}%
\def\MeV{\ifmmode {\mathrm{\ Me\kern -0.1em V}}\else
                   \textrm{Me\kern -0.1em V}\fi}%
\def\keV{\ifmmode {\mathrm{\ ke\kern -0.1em V}}\else
                   \textrm{ke\kern -0.1em V}\fi}%
\def\eV{\ifmmode  {\mathrm{\ e\kern -0.1em V}}\else
                   \textrm{e\kern -0.1em V}\fi}%
\let\tev=\TeV
\let\gev=\GeV
\let\mev=\MeV

\def\ifb{\mbox{fb$^{-1}$}}%  Inverse femtobarns.
\def\HERWIG{{{\sc Herwig} 6.520}\xspace}
\def\JIMMY{{{\sc Jimmy} 4.31}\xspace}
\def\PYTHIAEIGHT{{{\sc Pythia} 8.153}\xspace}
\def\PYTHIASIX{{{\sc Pythia} 6.426}\xspace}

\PreprintCoverPaperTitle{Search for type-III Seesaw heavy leptons in $pp$ collisions at $\sqrt{s}= 8 \tev$ with the ATLAS Detector}

\PreprintIdNumber{CERN-PH-EP-2015-094}

\PreprintJournalName{PRD}

\PreprintCoverAbstract{
A search for the pair-production  of heavy leptons ($N^0,L^{\pm}$) predicted by the type-III seesaw theory formulated to explain the origin of small neutrino masses is presented. The decay channels $N^0\rightarrow W^{\pm}l^{\mp}$  ($\ell = e, \mu, \tau$) and $L^{\pm}\rightarrow W^\pm \nu$ ($\nu=\nu_e, \nu_\mu, \nu_\tau$) are considered.
The analysis is performed using the final state that contains two leptons (electrons or muons),
two jets from a hadronically decaying $W$ boson, and large missing transverse momentum. The data used in the
measurement correspond to  an integrated luminosity of $20.3\,\ifb$ of $pp$ collisions at $\sqrt{s} = 8\,\tev$ 
collected by the ATLAS detector at the LHC. No evidence of heavy lepton pair-production is observed. 
Heavy leptons with masses below 325--540~GeV are excluded  at the 95\% confidence level, depending on the theoretical scenario considered.
}

\begin{document}

\title{Search for type-III Seesaw heavy leptons in $pp$ collisions at $\sqrt{s}= 8\,\tev$ with the ATLAS Detector}

\author{ATLAS Collaboration}

\begin{abstract}
A search for the pair-production  of heavy leptons ($N^0,L^{\pm}$) predicted by the type-III seesaw theory formulated to explain the origin of small neutrino masses is presented. 
The decay channels $N^0\rightarrow W^{\pm}l^{\mp}$  ($\ell = e, \mu, \tau$) and $L^{\pm}\rightarrow W^\pm \nu$ ($\nu=\nu_e, \nu_\mu, \nu_\tau$) are considered.
The analysis is performed using the final state that contains two leptons (electrons or muons),
two jets from a hadronically decaying $W$ boson and large missing transverse momentum. The data used in the
measurement correspond to  an integrated luminosity of $20.3\,\ifb$ of $pp$ collisions at $\sqrt{s} = 8\,\tev$ 
collected by the ATLAS detector at the LHC. No evidence of heavy lepton pair-production is observed. 
Heavy leptons with masses below 325--540~GeV are excluded at the 95\% confidence level, depending on the theoretical scenario considered.
\end{abstract}


\maketitle
\section{Introduction}
Experiments show that neutrinos have much smaller masses than charged leptons (see Ref. ~\cite{Agashe:2014kda}, and references therein).
While in the Standard Model (SM) the charged fermions acquire masses by coupling to the Higgs ($H$) boson, the neutrinos may become massive via new physics beyond the SM, e.g. via the introduction of Majorana mass terms~\cite{Ma:1998dn}. These masses could be small due to the seesaw mechanism~\cite{Minkowski:1977sc,Mohapatra:1979ia}, which relies on new massive states that couple to a charged lepton and the Higgs field.
Among different models for the seesaw mechanism, the type-III model~\cite{Foot:1988aq,Ma:1998dn}
introduces at least two extra triplets of fermionic fields with zero hypercharge in the adjoint representation of SU(2)$_{\rm L}$ that generate neutrino masses and couple to gauge bosons. 
This model predicts new charged and neutral heavy leptons that could be produced in proton-proton 
collisions at the Large Hadron Collider (LHC).

A search by the CMS experiment~\cite{CMS:2012ra} excluded
the type-III seesaw heavy leptons with masses  in the range
of 100-210 GeV, depending on theoretical assumptions. 
A recent search by ATLAS \cite{Aad:2015dha} also sets complementary limits on heavy leptons using the three-lepton final state.
Similar searches have also been done by L3 experiment~\cite{Achard:2001qw} 
ruling out charged heavy leptons with masses below 100 GeV.

In this paper, a search for heavy leptons predicted by the type-III seesaw mechanism is presented. The search explores the mass region above 100 GeV.
A minimal type-III seesaw model~\cite{Biggio:2011ja} is used to optimize the analysis strategy and interpret the search results. 
The model introduces a triplet with one neutral and two oppositely-charged leptons  
denoted by $N^0$ and $L^{\pm}$, respectively.
The heavy leptons decay into a SM lepton and a $W$, $Z$ or Higgs boson.
The heavy leptons are assumed to be degenerate in mass. This assumption does not affect the result because in the case of a small mass splitting due to radiative corrections, the decays within the heavy leptons are highly suppressed~\cite{Franceschini:2008pz}. 
The dominant production mechanism for type-III seesaw heavy leptons in $pp$ collisions is pair-production through the weak coupling to the $W$ boson propagator: $pp\rightarrow W^* \rightarrow  N^0L^{\pm}$, and the largest branching fraction
is the one with two $W$ bosons in the final state: $N^0\rightarrow W^{\pm}l^{\mp}$  ($\ell = e, \mu, \tau$) and $L^{\pm}\rightarrow W^\pm \nu$ ($\nu=\nu_e, \nu_\mu, \nu_\tau$).  
The production cross-section does not depend on the mixing angles between the SM leptons and the new heavy lepton states $V_\alpha, (\alpha=e,\mu,\tau$),  
which enter only in the expressions for the $L$ and $N$ decay widths. 
The fraction of $L$ and $N$ decays to lepton flavor $\alpha$ is proportional to $b_\alpha={|V_\alpha|^2}/({|V_e|^2 + |V_\mu|^2+|V_\tau|^2})$.
The limits obtained may be interpreted in terms of a range of mixing angles and Yukawa couplings~\cite{Biggio:2011ja}, allowing tests of a range of models with different couplings to gauge bosons and cross-section predictions~\cite{AguilarSaavedra:2009ik,Kumericki:2012bh}.
In the type-III seesaw model considered here, a benchmark point is defined by setting $V_\tau$ to zero, so that $b_e$ and $b_\mu$ are determined only by the ratio $V_e/V_\mu$, taken to be 0.87 based on the separately-allowed maximum values of $V_e$ and $V_\mu$ in Refs.~\cite{Abada:2008ea,Abada:2007ux,delAguila:2008pw}. This choice results in values of $b_e=0.53$, $b_\mu=0.4$3 and $b_\tau=0$. 
The search is performed for the process $pp\rightarrow N^0 L^{\pm}\rightarrow W^\pm\ell^\mp W^\pm\nu$,  
where one $W$ boson decays leptonically and the other $W$ boson decays hadronically, resulting in a lepton pair in the final state with either
the same charge (same-sign, SS) or with the opposite charge (opposite-sign, OS).

\section{Data sample and Monte Carlo simulation}
The analysis uses data from  $\sqrt{s}=8\TeV$ $pp$ 
collisions at the LHC that were recorded by the ATLAS detector using single-electron and single-muon triggers.
A detailed description of the  ATLAS detector can be found elsewhere~\cite{1748-0221-3-08-S08003}.
The data sample  corresponds to $20.3\pm0.6$\,\ifb~\cite{ATLAS_Luminosity2012} of integrated luminosity.
Data quality criteria are applied to ensure that  events were recorded with stable beam conditions and with 
all relevant subdetector systems operational. 
The triggers~\cite{Aad:2012xs} are fully efficient for leptons with $\pT>25$~GeV, where transverse momentum \pt\ is defined as the magnitude of the momentum component orthogonal to the beam axis.
Events are required to have a reconstructed collision vertex with at least three associated tracks, each with  $\pt>400\,\mev$.
In events with multiple vertices, the vertex with the largest $\sum\pt^2$  of associated tracks is taken as the primary event vertex.

Monte Carlo (MC) samples are used to optimize the event selection and
to model the kinematics and normalization of most background processes.
Signal samples are generated for heavy lepton masses in the range 100--600 $\GeV$.  
{\sc MadGraph 5}~\cite{Alwall:2011uj} is used to calculate the matrix elements for each process, while 
{\sc MadEvent}~\cite{Maltoni:2002qb} with the MSTW2008 parton distribution functions (PDF) set~\cite{Martin:2009iq} simulates the initial hard scattering and the $N$ and $L$ decay. \PYTHIAEIGHT~\cite{PYTHIA8} is used to simulate the decays 
of $W$  bosons and the underlying physics by providing parton showers and hadronization, 
as well as adding initial- and final-state radiation (ISR and FSR) to the events 
simulated in {\sc MadEvent}.  The main background sources arise from the production of a $Z$ boson in association with jets ($Z$+jets), 
single and pair-production of top quarks,  and diboson production ($WW, WZ, ZZ$). The $Z$+jets and diboson processes
are simulated with {\sc Sherpa} 1.4.1~\cite{Gleisberg:2008ta}, a generator based on a multileg matrix element calculation matched to the parton shower using the CKKW prescription~\cite{Catani:2001cc}, and using the CT10~\cite{PhysRevD.82.074024} PDF set.
For diboson production, both the electroweak and strong production processes are simulated~\footnote{Production 
mechanisms involving weak interactions at Born level (of order $\alpha^4_{\rm EW}$ without considering the boson 
decay, where $\alpha_{\rm EW}$ is the electroweak coupling constant) are referred to as electroweak production~\cite{Aad:2014zda}. 
Production mechanisms involving both the strong and electroweak interactions at Born level (of order 
$\alpha_{\rm S}^2\alpha^2_{\rm EW}$, where $\alpha_{\rm S}$ is the strong coupling constant) are referred to as strong 
production.}. Top-quark pair events and single-top-quark events in the $Wt$-channel and $s$-channel are simulated 
using MC@NLO 4.03~\cite{Frixione:2002ik}, which is interfaced to \HERWIG ~\cite{HERWIG_1} and 
\JIMMY ~\cite{Butterworth:1996zw} with the CT10 PDF set. 
Top pair production in association with a vector boson, $t\bar{t}+W/Z$,
is simulated using {\sc MadGraph} with the CTEQ6L1 PDF set~\cite{Pumplin:2002vw}, interfaced to \PYTHIAEIGHT\ for parton showering and hadronisation.
Single top quark production in the $t$-channel is simulated using {\sc AcerMC} 
v3.8~\cite{Kersevan:2004yg} with \PYTHIASIX~\cite{Sjostrand:2006za}  and the CTEQ6L1 PDF set.

All  samples of simulated events include the effect of multiple $pp$ interactions in the same and neighboring bunch crossings (pileup) by overlaying simulated minimum-bias events on each generated signal and background event. 
The number of overlaid events is chosen to match the average number of interactions per $pp$ bunch crossing observed in the data
as it evolved throughout the data-taking period (giving an average of 21 interactions per crossing for the whole data-taking period).
The generated samples are processed through the {\sc Geant4}-based detector simulation~\cite{Agostinelli:2002hh,:2010wqa} or a fast simulation using a parameterization of the performance of the calorimetry and {\sc Geant4} for the other parts of the detector~\cite{FastCaloSim}.
The standard ATLAS reconstruction software is used for both simulated and collision data.

\section{Object definitions}
% Leptons
The reconstructed objects used in this analysis are electrons, muons, jets, and missing transverse momentum. Electrons are reconstructed from clusters of energy depositions in the calorimeter that match a track reconstructed in the inner detector (ID) and satisfy the ``tight" criteria defined in Ref.~\cite{Aad:2014fxa}. The electrons are required to have $\pT > 25$ GeV and pseudorapidity $|\eta| < 2.47$~\footnote{ATLAS uses a right-handed coordinate system with its origin at the nominal interaction point (IP) in the center of the detector and the $z$-axis along the beam pipe. The $x$-axis points from the IP to the center of the LHC ring, and the $y$-axis points upward. Cylindrical coordinates $(r,\phi)$ are used in the transverse plane, $\phi$ being the azimuthal angle around the beam pipe. The pseudorapidity is defined in terms of the polar angle $\theta$ as $\eta=-\ln\tan(\theta/2)$}, 
excluding the transition region between the barrel and endcaps in the liquid argon calorimeter ($1.37 < |\eta| < 1.52$). 
Muons are reconstructed by combining ID and and muon spectrometer tracks that are spatially matched and have consistent curvatures.
The muon tracks are required to have $\pt>25$~GeV and $|\eta|<2.5$. In addition, leptons are required to be isolated from other tracks and calorimetric activity~\cite{Aad:2014hja}. 
To ensure that leptons originate from the interaction point, requirements of $|d_0|/\sigma_{d_0}<3$ and $|z_0 \sin\theta|<0.5$~mm are imposed on the electrons and muons, where $d_0$ ($z_0$) is the transverse (longitudinal) impact parameter of the lepton, and $\sigma_{d_0}$ is the uncertainty on the measured $d_0$. The lepton impact parameters are measured with respect to the event primary vertex. 

% AK4 jets
Jets are reconstructed from three-dimensional topological clusters of energy depositions in the calorimeter using the anti-$k_t$ algorithm~\cite{1126-6708-2008-04-063} with a radius parameter of $R=0.4$. 
 The energies of jets are calibrated to the hadronic energy scale  by correcting for energy losses in passive material, the non-compensating response of the calorimeter, and extra energy due to multiple $pp$ interactions~\cite{JES}. 
 The jets are required to have $\pT>30\,\GeV$ and $|\eta|<2.8$. For jets with $\pT<50\,\GeV$ and $|\eta|<2.4$, the summed scalar $\pT$ of associated tracks from the reconstructed primary vertex is required to be at least 25\% of the summed scalar $\pT$ of all tracks associated with the jet. In the pseudorapidity range $|\eta|<2.5$,
jets containing {\it b}-hadrons are identified using a $b$-tagging algorithm~\cite{b-tagging} with an efficiency of 70\% and with a misidentification rate for selecting light-quark or gluon jets of less than 1\%. The identification efficiency of the algorithm for jets containing {\it c}-hadrons is 20\%.
The efficiencies and misidentification rates are determined from $\ttbar$ MC events.

% MET
The missing transverse momentum vector (with its magnitude $\met$) is derived using the
calorimeter cell energies within $|\eta|<4.9$ and corrected on the basis of dedicated calibrations of the associated physics objects including
muons~\cite{ATLAS_MET}. Calorimeter cells containing energy depositions above noise and not associated with high-\pt\ physics objects are also included.

\section{Event selection}
% Event selection
Events that contain exactly two reconstructed leptons (electrons or muons), at least two jets, and no $b$-tagged jets are selected.
One of these leptons is required to match the object upon which the event was triggered. 
  Different sets of optimized selection criteria are 
used for the events in the OS and SS final states. 
The optimization is done using  simulated heavy lepton pair-production events at a benchmark mass of 300 GeV.
For the OS (SS) final  state, the leading and next-to-leading lepton candidates are required to have \pt\ greater than 
100 (70)\,\gev\ and  25 (40)\,\gev,  respectively.  The invariant mass of the two lepton 
candidates is required to be larger than  130 (90)\,\gev\ in order to suppress
background from the production of $Z$+jets. The hadronically decaying $W$ candidate is formed by combining the two jets with highest \pt, and 
the \pt\ of the first and second leading jets are respectively required to be larger than 60 (40) and 30 (25)\,\gev, for the OS (SS) 
final state.
The invariant mass of the $W$ candidate, $m_{jj}$, is required to be between 60 and $100\,\gev$. 
Events selected in the OS (SS) final state are required to have a \met\ of at least 110 (100)\,\gev\  and, for OS events, 
an angular separation $\Delta R_{jj}=\sqrt{(\Delta\eta)^2+(\Delta\phi)^2}<2$ between the two jets with highest 
\pt, where $\Delta\eta$ and $\Delta\phi$ are defined as the differences in pseudorapidity and azimuthal 
angle between the jets.

\begin{figure*}
  \begin{center}
    \includegraphics[width=0.48\textwidth] {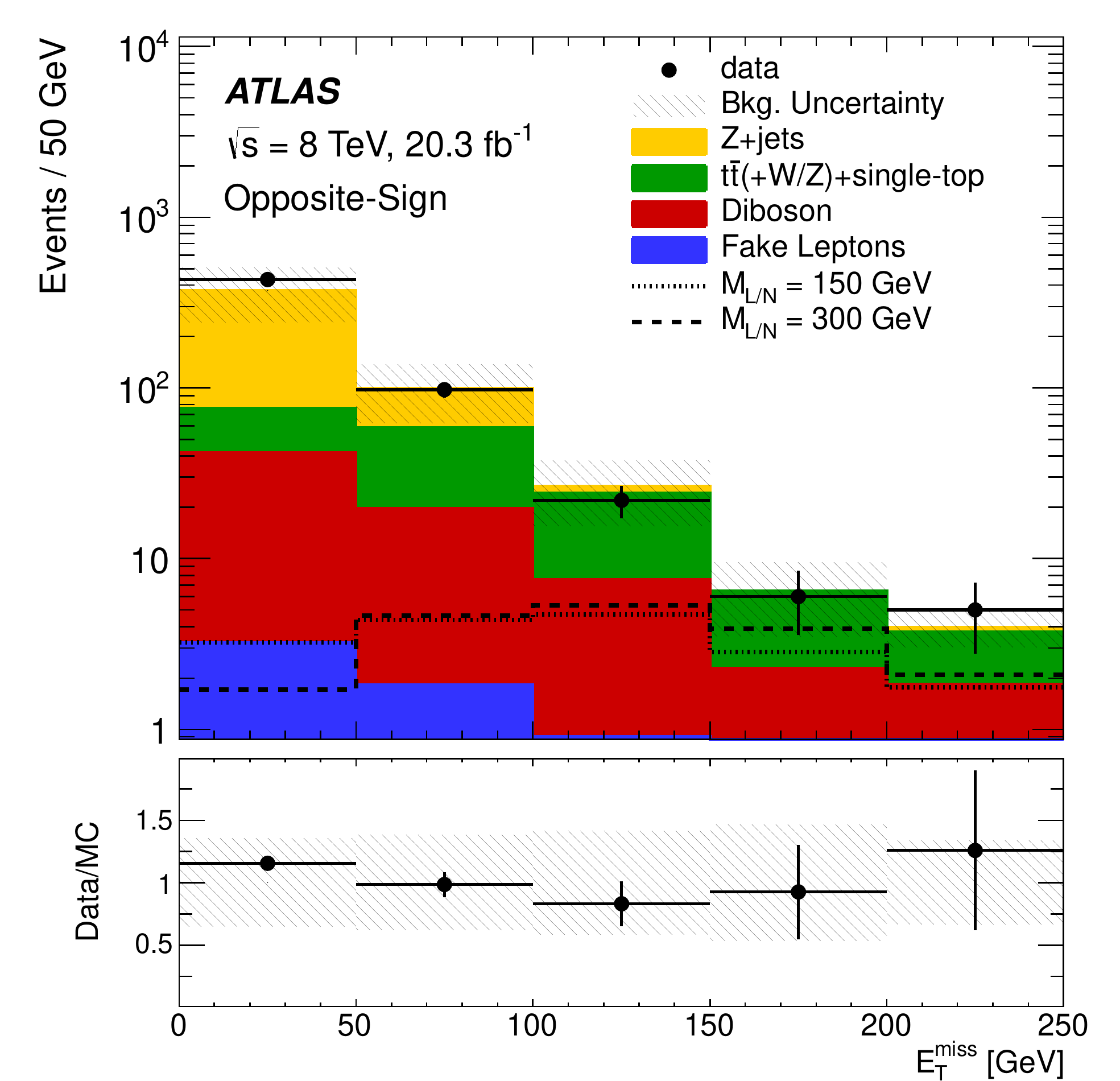}
    \includegraphics[width=0.48\textwidth] {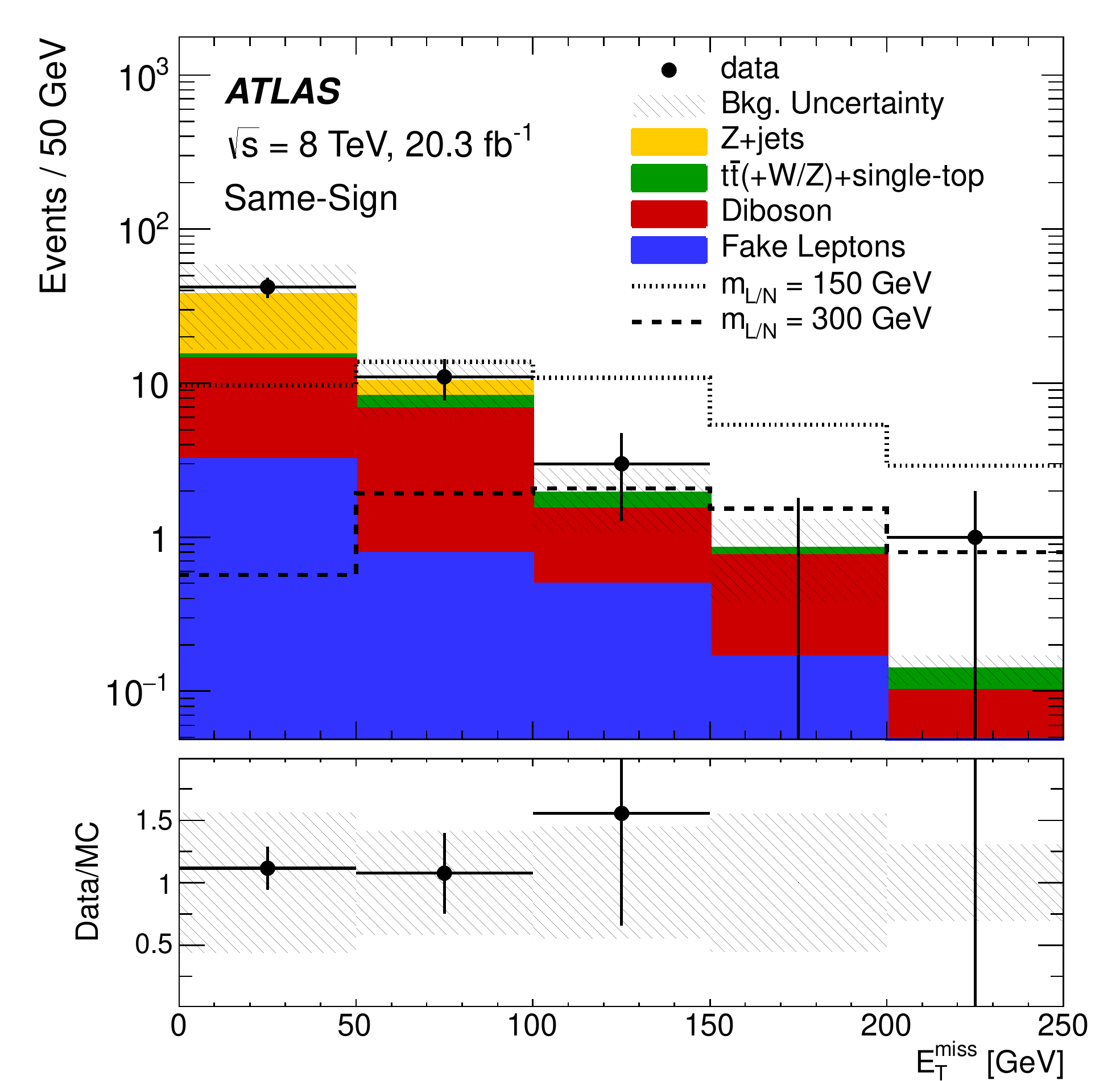}
    \caption{The missing transverse momentum distribution of opposite-sign (left) and same-sign (right) events for data and predictions. 
The dotted and dashed lines show the expected distribution of type-III seesaw lepton pair-production with masses of 150 and 300 \gev, respectively.
The events shown here are required to pass all selection requirements except that on the missing transverse momentum.
The uncertainties shown include both statistical and systematic uncertainties.
}
    \label{fig:metSignalRegion}
  \end{center}
\end{figure*}

\section{Background estimate}

The background in this search can be classified into two categories based on the origin of the charged lepton
candidates. The first category of backgrounds consists of events in which two leptons are 
produced via the decays of $W$ or $Z$ bosons and are correctly reconstructed. 
This category of backgrounds includes the production of $Z$+jets, $t\bar{t}$, $Wt$ single-top-quark and diboson events.
Smaller contributions originate from $t\bar{t}+W/Z$ events.
Contributions from triboson events, such as $WWW$,
and events containing a Higgs boson, are negligible. The number of events from this background category
is estimated using the  simulated samples described previously.

The second category corresponds to all other sources, such as events containing at least one particle that is incorrectly identified as a lepton,
or a lepton which originates from secondary interactions and decays, which are together denoted as {\it fakes}, 
and events with a lepton whose charge is incorrectly determined.
Fake electrons originate primarily from  jets that have a large electromagnetic energy fraction passing the
electron selection requirements,  photon conversions, and electrons from semileptonic decays of charm or bottom hadrons.
Fake muons include muons arising from semileptonic decays of charm or bottom hadrons,
in-flight decays of pions or kaons, or energetic particles that reach the muon spectrometer.

For the OS final state, the background contribution is dominated by events from the first category.
For the SS final state, the expected background is very small and is also primarily from the first category.
Events from $Z$+jets production can mimic the SS signal if the charge of one of the electrons is incorrectly determined.
The background events from this contribution are modeled by simulation, 
with correction factors derived using $Z\to ee$ events in data.
The probability of misidentifying the muon charge is negligible.

The background contribution in the second category (i.e. fake leptons) is  
estimated using an {\it in situ} technique~\cite{Aad:2014hja} that relies minimally on simulation.
This is done by reweighting a complementary set of events, selected by changing the electron identification criterion from ``tight" to ``loose"~\cite{ElectronID} and by loosening the muon $|d_0|/\sigma_{d_0}$ and the electron and muon isolation requirements, while keeping the event selection otherwise identical.
The reweighting factors are defined as the ratio of the number of events containing a lepton that satisfies the nominal criteria to the number of events containing a lepton that only fulfill the relaxed criteria.
These factors are measured as a function of the candidate \pt\ and $\eta$ in data samples
that are enriched in fake leptons~\cite{Aad:2014hja}. Corrections to the factors due to true
leptons from vector boson decay in the background-enriched samples are taken from MC simulation.

Figure~\ref{fig:metSignalRegion}  shows a comparison of the missing transverse momentum distribution of data, expected backgrounds and signal predictions
 when all the selection requirements, except for the missing transverse momentum requirement, are applied.
 The shape and the rate of the background estimate is in good agreement with the data.

\begin{figure}[hb!]
 % \begin{center}
    \includegraphics[width=1.\columnwidth] {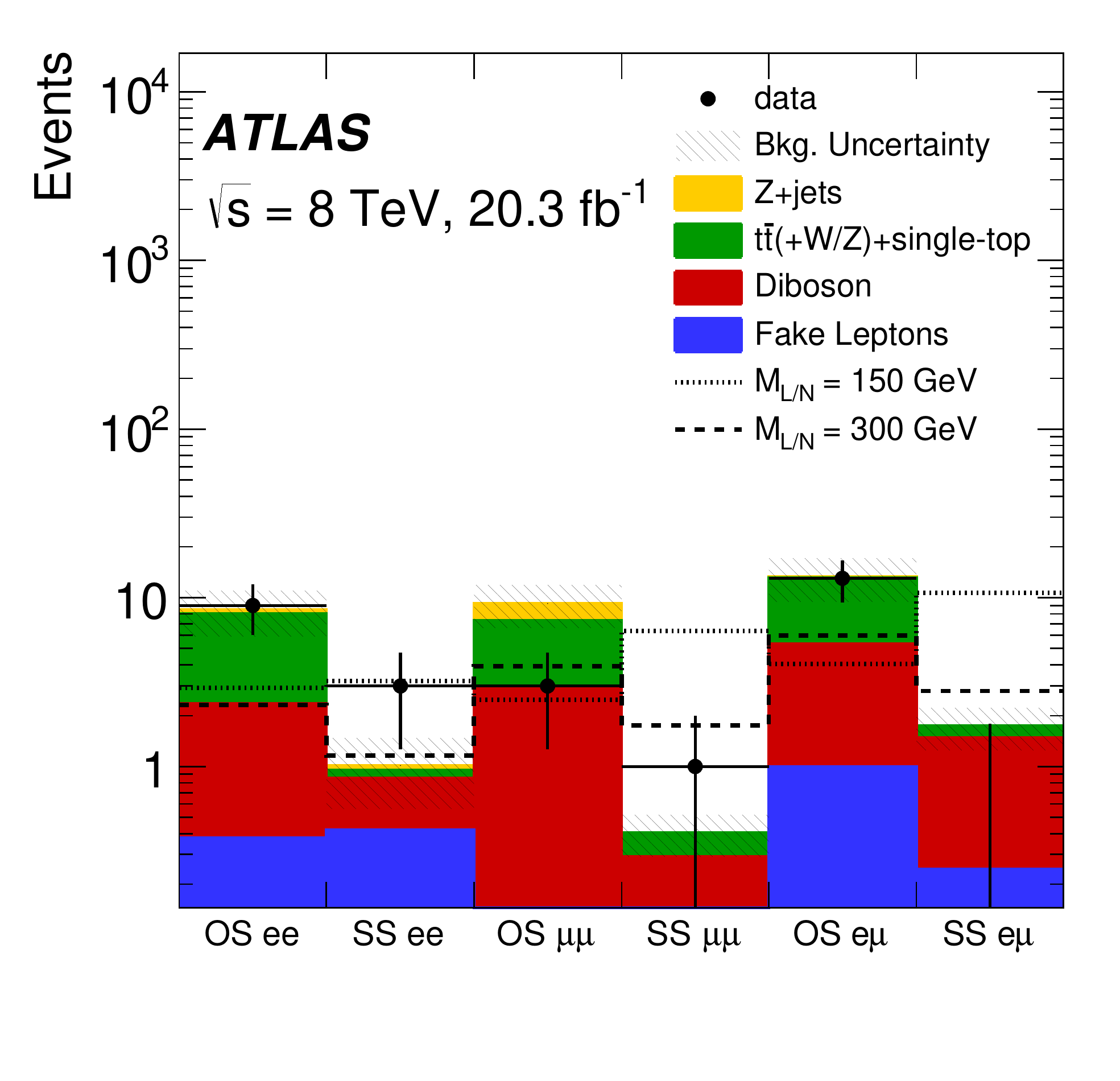}
    \caption{Event yields for opposite-sign (OS)  and  same-sign (SS) selection for
 $ee$, $\mu\mu$,  and $e\mu$  predicted backgrounds, data, and signal events featuring type-III seesaw lepton pair-production with masses 150 and 300 GeV.
The reported uncertainties include both the statistical and systematic uncertainties.
} \label{fig:flavComp}
 % \end{center}
\end{figure}

\begin{table}[b]
\begin{center}
\caption{Event yields for opposite-sign (OS)  and  same-sign (SS) selection for predicted backgrounds, data, and type-III seesaw lepton pair-production with masses of 150 and 300 \gev. The reported errors include both the statistical and systematic uncertainties.} \label{table:yields}
\begin{tabular}{lcc} \\\hline\hline
 & OS & SS \\\hline 
Fake Leptons 				& \numRP{1.36871}{1} $\pm$ \numRF{0.86430732}{1}  & \numRF{0.666395}{2}  $\pm$ \numRF{0.4247620}{2} \\ 
$Z$+jets 					& \numRF{2.44011}{2} $\pm$ \numRF{1.23384587}{2}  & \numRP{0.064125}{2}  $\pm$ \numRF{0.23051420}{2} \\ 
$WW/WZ/ZZ$ 				& \numRF{9.228404}{2} $\pm$ \numRF{2.87114107}{2}   & \numRP{1.9485481}{2} $\pm$ \numRF{0.58173979}{2}  \\ 
$t \bar{t}~(+W/Z) $ and single top 	& \numRP{17.91948}{1} $\pm$ \numRF{6.9472098131}{2}  & \numRF{0.4723918}{2} $\pm$ \numRF{0.250579727}{2}  \\ \hline
Total 					& \numRP{30.956704}{1} $\pm$ \numRF{7.66658842367}{2}  & \numRP{3.1514599}{2} $\pm$ \numRF{0.796725185022}{2}  \\ 
Data & $25$  & $4$\\
\hline 

Signal $m_{L/N}=150\gev$ & \numRF{9.45574}{2} $\pm$ \numRF{1.622142467}{2}  & \numRP{20.2903}{1}  $\pm$ \numRF{2.26406162}{2} \\
Signal $m_{L/N}=300\gev$ & \numRP{12.2103}{1} $\pm$ \numRF{0.62052860}{1}  & \numRP{5.69776}{1}  $\pm$ \numRF{0.54174661}{1} \\
\hline \hline
\end{tabular}

\end{center}
\end{table}

\begin{figure*}
 \includegraphics[width=0.99\columnwidth]{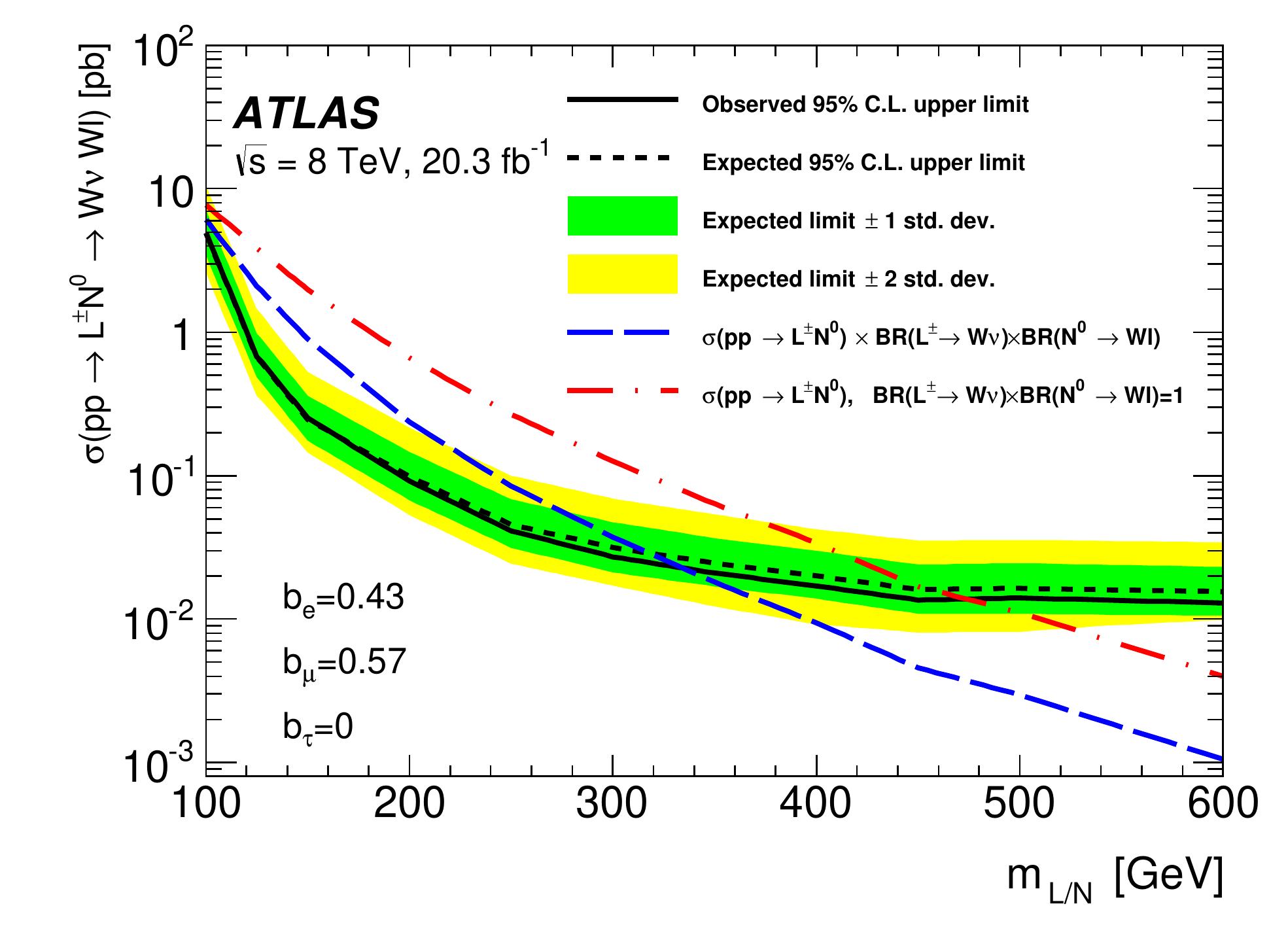}
\includegraphics[width=0.99\columnwidth]{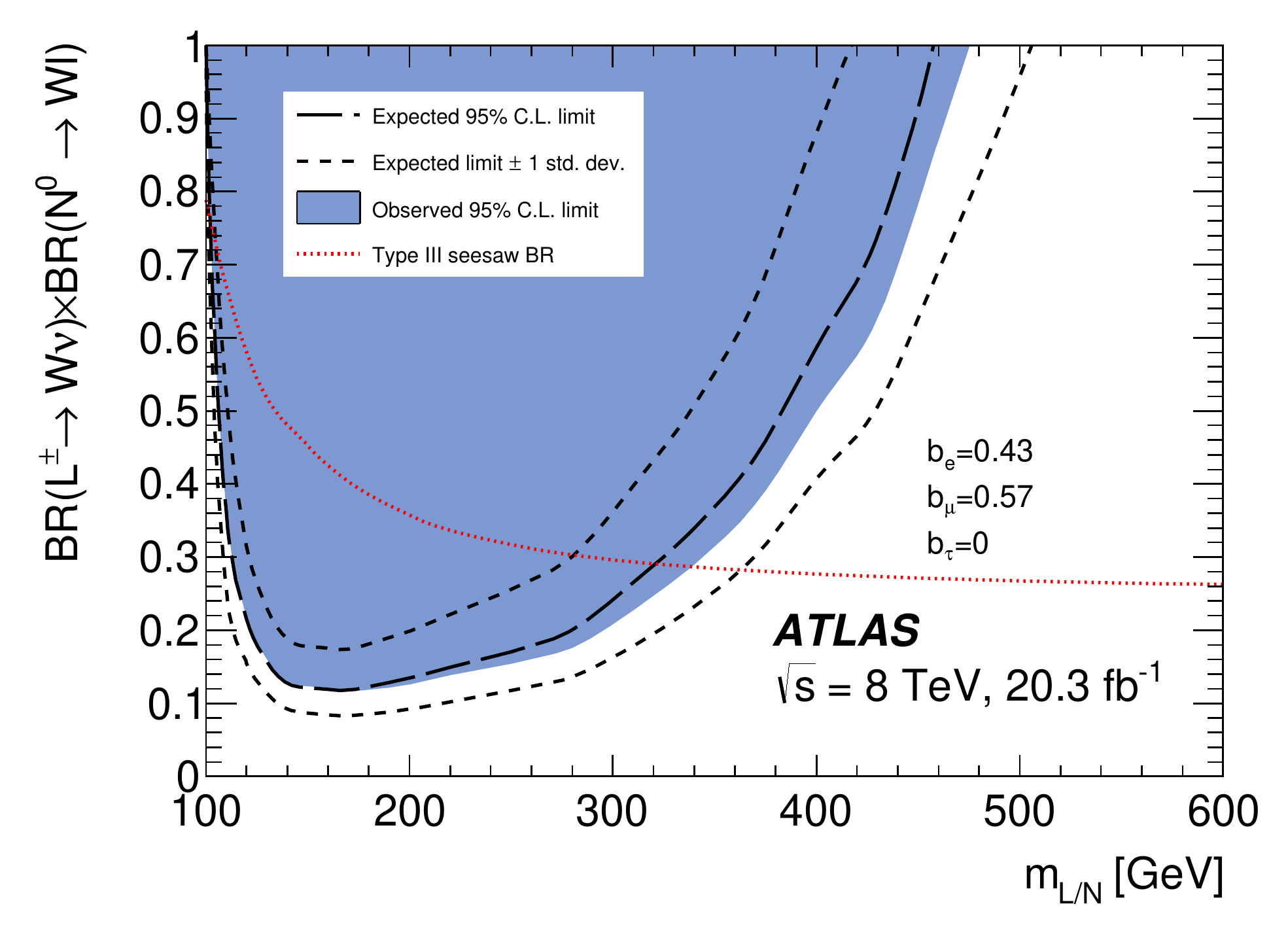}
\caption{Left: Observed (solid line) and expected (dashed line) 95\% C.L. upper limits on the type-III seesaw heavy lepton cross-section as a function of the  heavy lepton $L$ or $N$ mass assuming $b_{e}=0.43$, $b_{\mu}=0.57$, and $b_{\tau}=0$. The  bands surrounding the expected limit correspond to one and two standard deviations on the expected limit. The large-dashed  (dot-dashed) line  shows  the  theoretical prediction for $m_{L/N}$-dependent (maximal) branching fraction for decays to a $W$ boson. 
 Right: The expected (large-dashed line) and observed (shaded region) upper limits at the 95\% C.L.\ on the 
$BR(L^{\pm}\rightarrow W \nu) \times BR(N^0\rightarrow W^{\pm}l^{\mp})$ versus $m_{L/N}$. 
 The dashed line corresponds to one standard deviation around the expected limit. The dotted line shows the nominal mass-dependent branching ratio of the type-III seesaw model.}
\label{fig:limitStandard}
\end{figure*}

\begin{figure*}
\includegraphics[width=0.99\columnwidth]{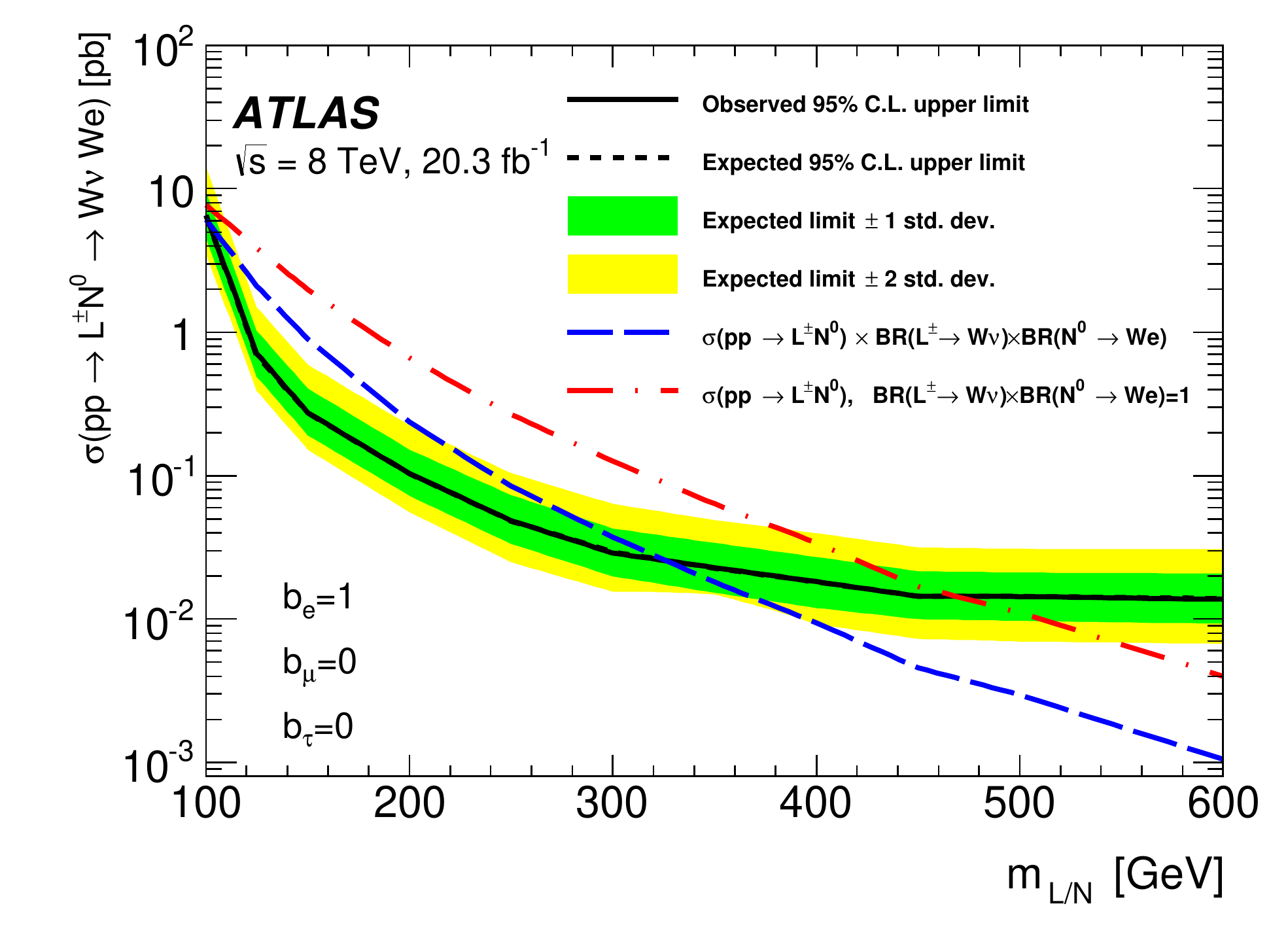}
\includegraphics[width=0.99\columnwidth]{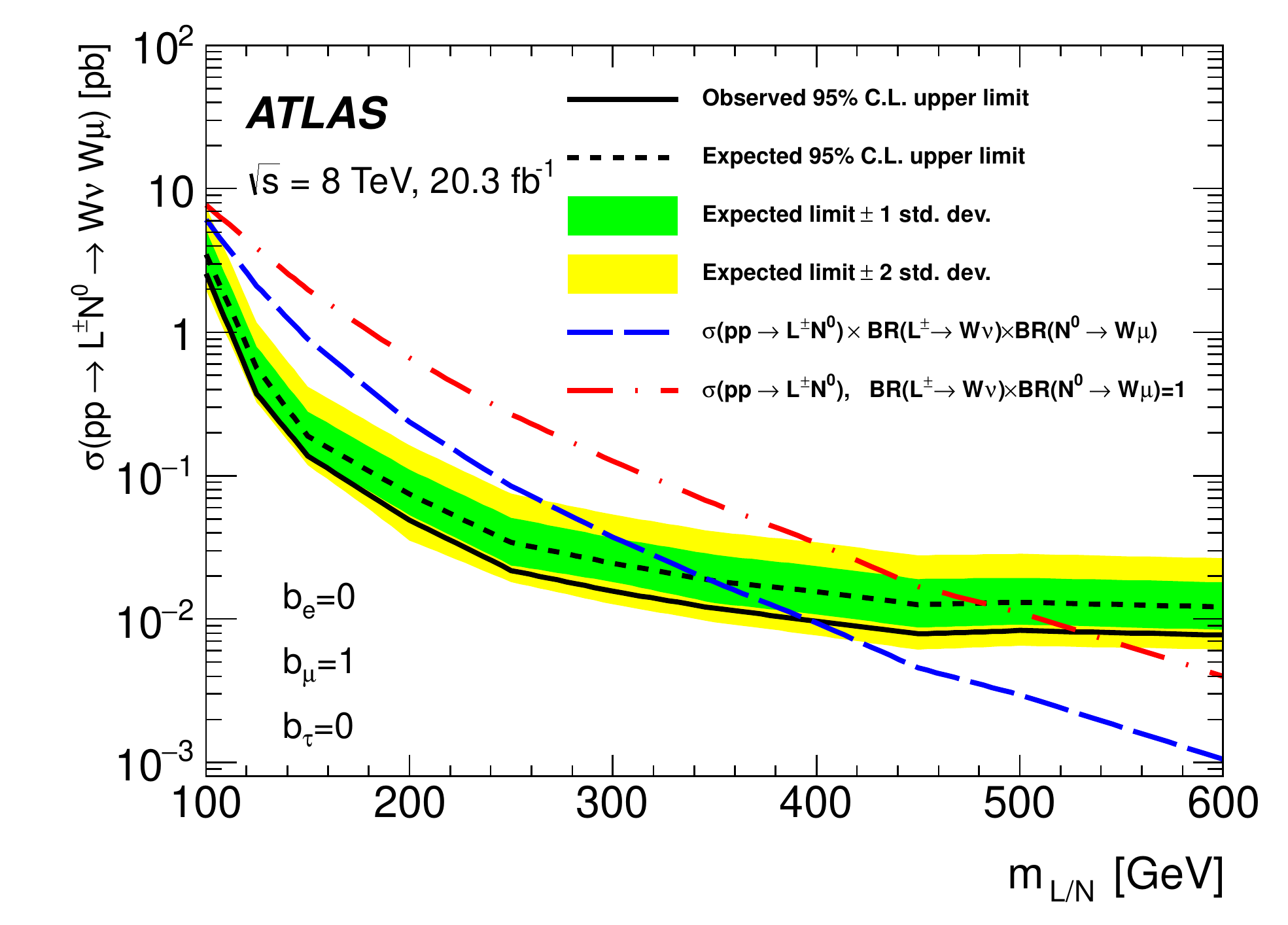}
\caption{Observed (solid line) and expected (dashed line) 95\% C.L.\ upper limits on the type-III seesaw heavy lepton cross-section as a function of the heavy lepton $L$ or $N$ mass for exclusive coupling to electrons (left) and exclusive coupling to muons (right). In the left plot, the dashed line is covered by the solid line. The large-dashed  (dot-dashed) line  shows  the  theoretical prediction for $m_{L/N}$  dependent (maximal) branching fraction for decays to a $W$ boson. 
}
\label{fig:limitCoupl}
\end{figure*}

The background estimates are validated by comparing the predicted numbers of events in simulation to those observed in data in several control regions that have event selection criteria similar to those for the signal region.

The control region for top quark pairs is defined by selecting events with two $b$-tagged jets.
In this region, according to MC simulation, all the events are from top quark pair and single top quark production with a negligible contribution from other sources. MC simulation predicts \numRF{26.2}{2}~$\pm$~\numRF{3.18}{1}~(stat.) events and in data 32 events are observed. The scale factor, the ratio of the observed and predicted event yields, is found to be consistent with unity.

The diboson control region is obtained using a $WZ$-enriched sample of events containing three leptons without a requirement on the missing transverse momentum. In this control region, according to MC simulation all the events are from diboson production with a negligible contribution from other sources. MC simulation predicts \numRF{10.5}{2} $\pm$ 1 events 
  and in data 9 events are observed.  The scale factor is found to be consistent with unity.

\section{Systematic uncertainties}
The uncertainties on the rate of top quark and diboson backgrounds
due to potential differences between data and MC simulation are evaluated using the statistical uncertainties of the measured scale factors in the control samples.
They are  the dominant systematic uncertainties ($\sim35\%$) on the background estimates.
For  top quark production, an additional systematic uncertainty is considered to account for
a potential difference between scale factors in the control region and the signal region.
This is done by comparing the nominal $t \bar{t}$ sample to alternative $t\bar{t}$ MC samples.
These samples include 
events that are generated using {\sc Powheg-Box } 1.0 (patch 4)\cite{Nason:2004rx,Frixione:2007vw,Alioli:2010xd} 
and the leading-order,  multileg generator {\sc Alpgen} v2.13 \cite{Mangano:2002ea}.
The {\sc Powheg-Box } generator is interfaced to the \PYTHIASIX showering routines with either CT10 or HeraPDF~\cite{Aaron:2009aa} PDF sets
and with the {\sc Powheg}  $h_{\rm damp}$ parameter set to either the mass of the top quark or infinity~\cite{ATL-PHYS-PUB-2013-005}.
{\sc Alpgen} is interfaced to \HERWIG, and used to simulate top pair events with up to four additional partons in the matrix element.
The uncertainties due to QCD ISR and FSR modeling are estimated with samples generated with {\sc AcerMC} v3.8 interfaced to \PYTHIASIX
in which the parton shower parameters are varied in a range consistent with a measurement of additional hadronic activity in $t\bar{t}$ events~\cite{ATLAS:2012al}.
The differences observed in the signal region by using different MC simulations are about 35\%.

For the $Z$+$n$-jets ($n\ge 2$) background estimate, the dominant systematic uncertainty in the OS final state is from the
uncertainty on its production cross-section ($\sim50\%$)~\cite{Alwall:2007fs}. For the SS final state,
the systematic uncertainty is dominated by the statistical uncertainty on the measured electron charge misidentification rate.

Uncertainties on the background estimate due to fake leptons are determined in dedicated studies using a combination of simulation and data. They
account for potential biases in the method used to extract the reweighting factors, and for the dependency of the reweighting factors on the event topology.

For both the predicted signal and background event yields,  uncertainties resulting from detector effects from jet energy scale and resolution~\cite{JES},
lepton reconstruction and identification efficiencies~\cite{Aad:2014fxa,Aad:2014rra}, lepton momentum scales and resolutions~\cite{Aad:2014nim,Aad:2014rra},
and missing transverse momentum~\cite{ATLAS_MET} are considered.  They are typically small (1--5\%).
The theoretical uncertainties on the  signal production cross-section and acceptance, such as PDF choice and  ISR and FSR modeling, are found to be negligible.

The background estimates and their uncertainties are tested in two other regions: a $Z$+jets control region and the hadronic $W$ sidebands.  
The $Z$+jets control region is selected by requiring the invariant mass of the oppositely-charged lepton pair to be consistent with the $Z$ boson mass.
In this region, approximately 70\% of the events are from a $Z$ boson produced in association with jets.
The predicted number of events is \numRF{33.718504}{2} $\pm$ \numRF{4.20981900517}{1}, where the error includes both the statistical and systematic uncertainties, and 32 events are observed in data. 
For the hadronic $W$ sidebands, the analysis was repeated using the same event selection, but requiring  the invariant mass
of the hadronically decaying $W$ candidate to be $35<m_{jj}<60\,\gev$ or 
$100<m_{jj}<125\,\gev$. This selection provides samples dominated by background events with kinematic properties similar to those of the signal candidates. 
In this region the predicted number of events is \numRF{33.87800}{2} $\pm$ \numRF{6.970950888356}{1} events, where the error includes both the statistical and systematic uncertainties, and in data 18 events are observed. The data are in agreement with the predictions within 1.9 standard deviations.

\section{Results and interpretation}
%%% Result/Interpretation
Table~\ref{table:yields} shows the predicted numbers of signal and background events and the observed data events in the signal region.
The data agree with the background-only hypothesis.
 Figure~\ref{fig:flavComp} shows the flavor composition of the simulated signal and background events, and of the observed events in data.
In the absence of any significant data excess, upper limits on the production rate of 
$pp\rightarrow N^0 L^{\pm}\rightarrow W^\pm\ell^\mp W^\pm\nu$ at the 95\% confidence level (C.L.) are derived as a function of the 
heavy lepton mass using the CL$_S$ method~\cite{Junk:1999kv}.
The results of limit calculations combining the observations in the OS and SS final states are shown in Figs.~\ref{fig:limitStandard}  and~\ref{fig:limitCoupl}.
With the default mixing angles considered here,  heavy leptons with masses less than $335\,\gev$ are
excluded by the analysis, while masses less than $475\,\gev$ are excluded  in the scenario   in which heavy leptons can only decay to the $W\ell$ or $W\nu$ final states.
For comparison, upper limits are also calculated for different theoretical assumptions,
such as exclusive coupling between the heavy leptons and muons ($b_e=0,b_{\mu}= 1$) or 
electrons ($b_e= 1,b_{\mu}=0$ ).
In the limit calculations for exclusive couplings, events in the two-muon (two-electron) final state are excluded 
for exclusive electron (muon) coupling. 
Heavy leptons with a mass below $400\, (325)\,\gev$ can be ruled out
by the data in the case of exclusive coupling to muons (electrons).
Masses less than $540\, (470) \,\gev$  are excluded in the scenario   in which heavy leptons can only decay to the $W\ell$ or $W\nu$ final state.
\section{Conclusions}
A search for the pair-production of heavy leptons predicted by the type-III seesaw model is presented. 
The analysis is performed using a final state that contains two leptons,
two jets from a hadronically decaying $W$ boson,  and large missing transverse momentum. The data used in the
search correspond to  an integrated luminosity of $20.3\,\ifb$ of $pp$ collisions at $\sqrt{s} = 8\,\tev$ 
collected by the ATLAS detector at the LHC. No evidence of heavy lepton production is observed. 
Heavy leptons with masses below 325--540~\gev\  are excluded at the 95\% confidence level, depending on the considered theoretical scenarios.

% Acknowledgements for papers with collision data
% Version 23-Mar-2015
% Standard acknowledgements start here

\section{Acknowledgments}

\begin{acknowledgments}
We thank CERN for the very successful operation of the LHC, as well as the
support staff from our institutions without whom ATLAS could not be
operated efficiently.

We acknowledge the support of ANPCyT, Argentina; YerPhI, Armenia; ARC,
Australia; BMWFW and FWF, Austria; ANAS, Azerbaijan; SSTC, Belarus; CNPq and FAPESP,
Brazil; NSERC, NRC and CFI, Canada; CERN; CONICYT, Chile; CAS, MOST and NSFC,
China; COLCIENCIAS, Colombia; MSMT CR, MPO CR and VSC CR, Czech Republic;
DNRF, DNSRC and Lundbeck Foundation, Denmark; EPLANET, ERC and NSRF, European Union;
IN2P3-CNRS, CEA-DSM/IRFU, France; GNSF, Georgia; BMBF, DFG, HGF, MPG and AvH
Foundation, Germany; GSRT and NSRF, Greece; RGC, Hong Kong SAR, China; ISF, MINERVA, GIF, I-CORE and Benoziyo Center, Israel; INFN, Italy; MEXT and JSPS, Japan; CNRST, Morocco; FOM and NWO, Netherlands; BRF and RCN, Norway; MNiSW and NCN, Poland; GRICES and FCT, Portugal; MNE/IFA, Romania; MES of Russia and NRC KI, Russian Federation; JINR; MSTD,
Serbia; MSSR, Slovakia; ARRS and MIZ\v{S}, Slovenia; DST/NRF, South Africa;
MINECO, Spain; SRC and Wallenberg Foundation, Sweden; SER, SNSF and Cantons of
Bern and Geneva, Switzerland; NSC, Taiwan; TAEK, Turkey; STFC, the Royal
Society and Leverhulme Trust, United Kingdom; DOE and NSF, United States of
America.

The crucial computing support from all WLCG partners is acknowledged
gratefully, in particular from CERN and the ATLAS Tier-1 facilities at
TRIUMF (Canada), NDGF (Denmark, Norway, Sweden), CC-IN2P3 (France),
KIT/GridKA (Germany), INFN-CNAF (Italy), NL-T1 (Netherlands), PIC (Spain),
ASGC (Taiwan), RAL (UK) and BNL (USA) and in the Tier-2 facilities
worldwide.

\end{acknowledgments}

%\bibliography{paper_heavylepton}

%

\onecolumngrid

\clearpage

% ATLAS Collaboration author list
% Data extracted on 11-Aug-2015 for paper reference EXOT-2014-07
%\documentclass[11pt]{article}
%\usepackage{a4wide}\begin{document}
\begin{flushleft}
{\Large The ATLAS Collaboration}

\bigskip

G.~Aad$^{\rm 85}$,
B.~Abbott$^{\rm 113}$,
J.~Abdallah$^{\rm 151}$,
O.~Abdinov$^{\rm 11}$,
R.~Aben$^{\rm 107}$,
M.~Abolins$^{\rm 90}$,
O.S.~AbouZeid$^{\rm 158}$,
H.~Abramowicz$^{\rm 153}$,
H.~Abreu$^{\rm 152}$,
R.~Abreu$^{\rm 30}$
% ...
\bigskip
\\
% ...
$^{11}$ Institute of Physics, Azerbaijan Academy of Sciences, Baku, Azerbaijan\\
% ...
$^{30}$ CERN, Geneva, Switzerland\\
% ...
$^{85}$ CPPM, Aix-Marseille Universit{\'e} and CNRS/IN2P3, Marseille, France\\
% ...
$^{90}$ Department of Physics and Astronomy, Michigan State University, East Lansing MI, United States of America\\
% ...
$^{107}$ Nikhef National Institute for Subatomic Physics and University of Amsterdam, Amsterdam, Netherlands\\
% ...
$^{113}$ Homer L. Dodge Department of Physics and Astronomy, University of Oklahoma, Norman OK, United States of America\\
% ...
$^{151}$ Institute of Physics, Academia Sinica, Taipei, Taiwan\\
$^{152}$ Department of Physics, Technion: Israel Institute of Technology, Haifa, Israel\\
$^{153}$ Raymond and Beverly Sackler School of Physics and Astronomy, Tel Aviv University, Tel Aviv, Israel\\
% ...
$^{158}$ Department of Physics, University of Toronto, Toronto ON, Canada\\
% ...
\end{flushleft}

\end{document}
%
% ****** End of file apstemplate.tex ******